\documentclass[aps,prd,onecolumn,showpacs,amsmath,amssymb]{revtex4}
\usepackage{graphicx}
\usepackage{dcolumn}
\usepackage{bm}
\def\beqra{\begin{eqnarray}}
\def\eeqra{\end{eqnarray}}
\def\beq{\begin{equation}}
\def\eeq{\end{equation}}

\begin{document}
\newcommand{\Spin}[4]{\, {}_{#2}^{\vphantom{#4}} {#1}_{#3}^{#4}}
\newcommand{\vertsp}{\vphantom{\displaystyle{\dot a \over a}}}
\newcommand{\Gm}[3]{\, {}_{#1}^{\vphantom{#3}} G_{#2}^{#3}}
\newcommand{\Spy}[3]{\, {}_{#1}^{\vphantom{#3}} Y_{#2}^{#3}}
\draft

\title{CMB polarization from secondary vector and tensor modes}

\author{Silvia Mollerach}
\affiliation{Centro At\'omico Bariloche, Av. Bustillo 9500 \\ 8400
Bariloche, Argentina}
\email{mollerach@cab.cnea.gov.ar}
\author{Diego Harari}
\affiliation{Departamento de F\'{i}sica, FCEyN, Universidad de Buenos Aires \\
Ciudad Universitaria - Pab. 1, 1428, Buenos Aires, Argentina}
\email{harari@df.uba.ar}
\author{Sabino Matarrese}
\affiliation{Dipartimento di Fisica `G. Galilei',
Universit\`{a} di Padova \\ INFN -
Sezione di Padova \\ via Marzolo 8, I-35131 Padova, Italy}
\email{matarrese@pd.infn.it}

\date{\today}

\begin{abstract}

We consider a novel contribution to the polarization of the Cosmic Microwave 
Background induced by vector and tensor modes generated by the non-linear 
evolution of primordial scalar perturbations. Our calculation is based on 
relativistic second-order perturbation theory and allows to estimate the 
effects of these secondary modes on the polarization angular power-spectra. 
We show that a non-vanishing B-mode polarization unavoidably arises
from pure scalar initial perturbations, thus limiting our ability 
to detect the signature of primordial gravitational waves 
generated during inflation. This secondary effect dominates over 
that of primordial tensors for an inflationary tensor-to-scalar ratio 
$r<10^{-6}$. The magnitude of the effect is smaller than the
contamination produced by the conversion of polarization  of type E into type 
B, by weak gravitational lensing. However the lensing signal can be
cleaned, making the secondary modes discussed here the actual
background limiting the detection of small amplitude primordial
gravitational waves. 

\end{abstract}

\maketitle

\section{Introduction}

The generation of a stochastic background of primordial gravitational 
waves is a fundamental prediction of inflationary models 
for the early Universe. Its amplitude is determined by the energy scale of 
inflation, which can widely vary 
between different inflationary models. The detection of this gravitational 
radiation would provide a crucial test for the validity of the whole 
scenario.  
Gravitational wave detectors, however, are quite unlikely to have enough
sensitivity to detect such a primordial signal, owing both to its
smallness and to its extremely low characteristic frequencies.
The existence of ultra-low-frequency gravitational radiation, however, 
can be indirectly probed thanks to the temperature anisotropy and 
polarization it induces on the Cosmic Microwave Background (CMB) 
radiation. In particular, the curl component, called B-mode, of the 
CMB polarization provides a unique opportunity to disentangle the 
effect of tensor (gravitational-wave) from scalar 
perturbations, as this is only excited by either tensor or 
vector modes \cite{se97,ka97}. From this point of view, future satellite 
missions, such as {\em Planck}, which will have enough sensitivity to either 
detect or constrain the B-mode CMB polarization predicted by the simplest 
inflationary models, might represent the first `space-based gravitational-wave 
detector' \cite{ckw}. 
The main complication in this context comes from the effect of 
gravitational lensing on the CMB by the matter distribution, which
implies the transformation of E-mode into B-mode polarization 
\cite{zalsel98}: such a non-linear effect might actually obscure the signal 
due to primordial tensor modes. It has been pointed out that the inflationary 
gravitational-wave background can only be detected by CMB polarization 
measurements if the tensor to scalar ratio $r \ge
10^{-4}$, which corresponds to an energy scale of inflation 
larger than $3\times 10^{15}$ GeV \cite{knox,kesden,kesden03,kinney}.  
Quite recently, however, a better technique to {\it clean} polarization maps
from the lensing effect has been proposed, which would allow tensor-to-scalar 
ratios as low as $10^{-6}$, or even smaller, to be probed 
\cite{hirata,seljak03}. 
Other secondary contributions to the B-type polarization arising
during the reionization stage, though of much smaller amplitude,
have been considered in \cite{hu00}.

The case of vector modes is even more 
interesting, as they cannot be produced during 
inflation and are in general extremely difficult to generate at 
early epochs, with the exception of models which predict the existence 
of primordial tangled magnetic fields \cite{seshadri,subramanian}. 

This paper considers a new source of B-mode polarization,  
coming from secondary vector and tensor modes. The contribution to 
temperature anisotropy arising from these modes has already been analyzed in 
Refs. \cite{mm97,mol97}. 
The evolution of cosmological perturbations away from the linear regime
is in fact characterized by mode-mixing, which not only implies that different 
Fourier modes influence each other, but also that primordial density 
perturbations act as a source for curl vector perturbations and gravitational 
waves \footnote{Also, primordial gravitational waves give rise to 
second-order scalar and vector perturbations, but this effect is usually 
negligible \cite{mmb}.} \cite{mmb}. 
Let us emphasize that these secondary vector and tensor modes always exist 
and that their amplitude has a one-to-one relation with the level of 
density perturbations, which is severely constrained by both CMB anisotropy 
measurements and Large-Scale Structure observations. 
Therefore, their properties are largely inflation model-independent, 
contrary to primary tensor modes whose amplitude is not only 
model-dependent, but is well-known to be suppressed in some cases, like e.g. 
in the so-called {\em curvaton} model for the generation of 
curvature perturbations \cite{curvaton}.

The plan of the paper is as follows. In Section II we introduce 
the second-order vector and tensor modes which are produced by the non-linear 
evolution of primordial scalar perturbations. 
In Section III we obtain the contribution of these secondary modes 
to the polarization angular power-spectra, while in Section IV we compare 
these contributions to those from primordial gravitational waves 
and gravitational lensing. Section V contains our main conclusions.

\section{Second-order vector and tensor modes}

The perturbed line-element around a spatially flat 
FRW background takes a particularly simple form in the 
so-called  Poisson gauge \cite{maber}, which to linear order reduces to 
the Newtonian gauge. Adopting the conformal time $d\eta=dt/a$, one can write
\begin{equation}
\label{linelem}
d s^2 = a^2(\eta)\left\{-\left(1+2\Psi\right)d\eta^2 
- 2 V_i d\eta dx^i + \left[\left(1-2\Phi\right)\delta_{ij} + 
2 H_{ij}\right]dx^i dx^j \right\} \;, 
\end{equation}
with $\delta_{ij}$ the Kronecker symbol. 
The lapse perturbation is the sum of a first-order or primary term
(indicated by a label P) and a second-order one
(indicated by a label S): $\Psi=\Psi_{\rm P} + \Psi_{\rm S}$. 

The shift perturbation $V_i$ in this gauge is a pure vector,
i.e. $\partial^i V_i=0$, which only arises as a second (or higher)-order 
contribution (i.e. $V_i = V_{{\rm S}i}$).  
The spatial metric perturbations contain a scalar mode, which includes 
both a linear and a second-order term, 
$\Phi = \Phi_{\rm P} + \Phi_{\rm S}$,
and a tensor (i.e. transverse and traceless) mode  
$H_{ij}$ ($\partial^i H_{ij}= H^i_{~i}= 0$), which also 
contains first and second-order contributions, namely
$H_{ij} = H_{{\rm P}ij} + H_{{\rm S}ij}$. 
Here and in what follows spatial indices are raised by the
Kronecker symbol $\delta^i_j$. 

Hereafter we assume that the Universe is spatially flat and filled with a 
cosmological constant $\Lambda$ and a pressureless fluid -- made of Cold Dark 
Matter (CDM) plus baryons -- whose stress energy-tensor reads 
$T^\mu_{~\nu}=\rho ~u^\mu u_\nu$ ($u^\mu u_\mu =-1$). 
This is a reasonable aproximation for the purposes of this paper,
although the inclusion of the radiation in the evolution equations
would give more accurate results at the smaller scales considered,
that entered the horizon during the radiation dominated era.

In this paper we will concentrate on vector and tensor 
perturbations. 
In the standard inflationary scenario for the origin of perturbations
linear vector modes are not present, but they are generated 
after horizon crossing as non-linear combinations of primordial scalar modes. 

Let us report here the second-order calculation which leads to 
the generation of these secondary vector and tensor modes. 
For the $\Lambda=0$ case, this problem was originally solved in Ref. 
\cite{mmb} starting from the results of second-order calculations in the 
synchronous gauge \cite{tomita,mps} and transforming them to the Poisson 
gauge, by means of second-order gauge transformations \cite{bmms}. 
Here we will solve the problem directly in the Poisson gauge and we will 
account for a non-vanishing $\Lambda$ term (see also \cite{bruni}).  

The background Friedmann equations read 
$3{\cal H}^2=  a^2 \left(8\pi G \bar\rho + \Lambda\right)$ and
$\dot{\bar\rho} = - 3 {\cal H} \bar\rho$, where dots 
indicate differentiation with respect to $\eta$, 
${\cal H} \equiv \dot a /a$ and $\bar\rho$ is the 
mean mass-density.  Also useful is the relation
$2\dot{\cal H} = -{\cal H}^2 + a^2 \Lambda$.

We then perturb the mass-density and fluid four-velocity as 
$\rho =\bar\rho (1 + \delta)$ and
$u^\mu =(\delta^\mu_0 + v^\mu)/a$,
with $\delta = \delta_{\rm P} + \delta_{\rm S}$
and $v^\mu = v_{\rm P}^\mu + v_{\rm S}^\mu$ \cite{mmb}.

Let us briefly report the results of linear perturbation theory in this gauge 
(for a detailed analysis, see e.g. \cite{fmb}). The non-diagonal 
components of $i$-$j$ Einstein's equations imply $\Psi_{\rm P}
= \Phi_{\rm P} \equiv \varphi$, for the scalar part, and
\begin{equation}
\label{gravwaves}
\ddot{H}^i_{{\rm P}j} + 2{\cal H} \dot{H}^i_{{\rm P}j}
- \nabla^2  H^i_{{\rm P}j} =0 \;, 
\end{equation}
for the tensor part, 
while its trace gives the evolution equation for the linear 
scalar potential $\varphi$, namely 
\begin{equation}
\label{master}
\ddot \varphi + 3 {\cal H} 
\dot \varphi + a^2 \Lambda \varphi = 0 \;.
\end{equation}
Selecting only the growing-mode solution, we can write
$\varphi({\bf x},\eta) = \varphi_0({\bf x}) g(\eta)$, where 
$\varphi_0$ is the peculiar gravitational
potential linearly extrapolated to the present time ($\eta_0$) and  
$g=D_+/a$ is the so-called 
growth-suppression factor, where $D_+(\eta)$ is the linear growing-mode of 
density fluctuations in the Newtonian limit and $a$ the scale-factor. 
In the $\Lambda=0$ case $g=1$. An excellent approximation for $g$ as a 
function of redshift $z$ is given in 
Refs. \cite{lahav,cpt} 
\begin{equation}
g \propto \Omega_m\left[\Omega_m^{4/7} - \Omega_\Lambda +
\left(1+ \Omega_m/2\right)\left(1+ \Omega_\Lambda/70\right)\right]^{-1} \;, 
\end{equation}
with $\Omega_m=\Omega_{0m}(1+z)^3/E^2(z)$, 
$\Omega_\Lambda=\Omega_{0\Lambda}/E^2(z)$, 
$E(z) \equiv \left[\Omega_{0m}(1+z)^3 + 
\Omega_{0\Lambda}\right]^{1/2}$ and 
$\Omega_{0m}$, $\Omega_{0\Lambda}=1-\Omega_{0m}$, the present-day
density parameters of non-relativistic matter and cosmological constant, 
respectively. We will normalize the growth-suppression factor so that 
$g(z=0)=1$. 

The energy and momentum constraints provide the density and
velocity fluctuations in terms of $\varphi$, namely
\begin{eqnarray} 
\label{lindensvel}
\delta_{\rm P} & = & \frac{1}{4 \pi G a^2 \bar\rho} 
\left[ \nabla^2 \varphi - 3{\cal H}\left(\dot\varphi + {\cal H} 
\varphi\right) \right],
\nonumber \\
v_{{\rm P}i} & = & - \frac{1}{4 \pi G a^2 \bar\rho} \partial_i 
\left(\dot\varphi + {\cal H} \varphi\right) \;.  
\end{eqnarray}

Perturbations of the matter stress-energy tensor up to second-order 
in the Poisson gauge have been calculated in Ref. \cite{bmr03}, for a 
general perfect fluid. Specializing to the pressureless case, one has 
\begin{eqnarray}
T^0_{~0} & = & -\bar\rho \left(1+ \delta_{\rm P} + \delta_{\rm S} + 
v_{\rm P}^2 \right) \;, \nonumber \\
T^i_{~0} & = &  -\bar\rho \left[ v_{\rm P}^i + v_{\rm S}^i
+\left(\varphi+\delta_{\rm P}\right)v_{\rm P}^i \right]   \;, \nonumber \\
T^0_{~i} & = &  \bar\rho \left[ v_{{\rm P}i} + v_{{\rm S}i}
- V_i 
+\left(-3\varphi+\delta_{\rm P}\right)v_{{\rm P}i} \right] \;, \nonumber \\
T^i_{~j} & = &  \bar\rho ~ v_{\rm P}^i v_{{\rm P}j} \;,
\end{eqnarray}
where $v_{\rm P}^2 \equiv v_{\rm P}^j v_{{\rm P}j}$. 
Note that the second-order velocity $v_{{\rm S}i}$ is the sum of an 
irrotational component $v^{(0)}_{{\rm S}i}$, which is the gradient of a 
scalar, and a rotational vector $v^{(1)}_{{\rm S}i} $, 
which has zero divergence, $\partial^i v^{(1)}_{{\rm S}i}=0$. 

For the purpose of obtaining secondary vector and tensor modes, 
we can start by writing the second-order momentum-conservation 
equation \footnote{Second-order Christoffel symbols can be found in  
Appendix A of Ref.\cite{abmr} (see also \cite {noh}).}, 
$T^\mu_{~i;\mu}=0$, which gives
\begin{equation}
\left(\dot{v}^{(1)}_{{\rm S}i} - \dot{V}_i \right) + 
{\cal H} \left(v^{(1)}_{{\rm S}i} - V_i \right) =
- \partial_i \Psi_{\rm S} - 2 \dot\varphi v_{{\rm P}i} 
- \frac{1}{2} \partial_i \left(v_{\rm P}^2 + \varphi^2 \right) \;.  
\end{equation}
For pure growing-mode initial conditions $\dot\varphi \propto \varphi$ 
and $v_{{\rm P}i} \propto \partial_i \varphi$, which makes the RHS of this
equation the gradient of a scalar quantity; thus, the vector part only  
contains a decaying solution $(v^{(1)}_{{\rm S}i} -V_i) 
\propto a^{-1}$, and we can safely assume 
$v^{(1)}_{{\rm S}i} = V_i$. 

To proceed one needs the second-order perturbations of 
the Einstein tensor, $\delta^{(2)}G^\mu_{~\nu}$: these can be found 
for any gauge in Appendix A of Ref. \cite{abmr}.  

The second-order `momentum constraint' $\delta^{(2)}G^i_{~0} 
= 8 \pi G \delta^{(2)}T^i_{~0}$ gives 
\begin{equation}
\partial^i \left( {\cal H} \Psi_{\rm S} +  \dot\Phi_{\rm S} \right) 
+\frac{1}{4} \nabla^2 V^i
+ \dot\varphi \partial^i \varphi 
+ 4 \varphi \partial^i \dot\varphi
 = - 4 \pi G a^2 \bar\rho \left[\left(\varphi + 
\delta_{\rm P}\right) v_{\rm P}^i + v_{\rm S}^{(0)i}\right]\;. 
\end{equation}

The pure vector part of this equation can be isolated by first taking its 
divergence to solve for the combination ${\cal H}\Psi_{\rm S} + 
\dot\Phi_{\rm S}$ and then replacing it in the original equation.
We obtain 
\begin{equation}
\nabla^2 \nabla^2 V_i  = 16 \pi G a^2 \bar\rho 
~\partial^j \left(  v_{{\rm P}j} \partial_i \delta_{\rm P} -  
v_{{\rm P}i} \partial_j \delta_{\rm P} \right)\; . 
\end{equation}

We can further simplify this equation and write: 
\begin{equation}
\label{eq:vectors}
\nabla^2  V_i = - \frac{8}{3}F(z)
\left(\partial_i\varphi_0 \nabla^2\varphi_0
- \partial^i\partial^j \varphi_0 \partial_j\varphi_0 + 
2 \partial_j\Theta_0 \right) \;, 
\end{equation}
with 
\begin{equation}
F(z)=\frac{2 g^2(z) E(z)
f(\Omega_m)} {\Omega_{0m} H_0 \left(1+z\right)^2}\;, 
\nonumber
\end{equation}
where \cite{lahav,cpt} $f(\Omega_m) \equiv d \ln D_+/d \ln a
\approx \Omega_m(z)^{4/7}$, $H_0$ is the Hubble constant and 
\begin{equation}
\label{eq:Theta}
\nabla^2\Theta_0 = -\frac{1}{2}\left((\nabla^2\varphi_0)^2-
\partial_i \partial_k \varphi_0 \partial^i \partial^k 
\varphi_0\right) \;.
\end{equation}

For $\Lambda=0$, the above expression for $V_i$ 
reduces to Eq. (6.8) of Ref. \cite{mmb}, noting that $F(z)=\eta$ in
that case. 

Finally, we can solve for the second-order tensor modes by
looking at the traceless part of $i$-$j$ Einstein's equations, 
$\delta^{(2)}G^i_{~j} - \frac{1}{3}\delta^{(2)}G^k_{~k} 
\delta^i_j = 8 \pi G (\delta^{(2)}T^i_{~j} - 
\frac{1}{3} \delta^{(2)}T^k_{~k} \delta^i_j)$, namely 
\begin{eqnarray}
&& - \left[\frac{1}{3}\nabla^2\left(\Phi_{\rm S} - \Psi_{\rm S}\right) + 
\frac{2}{3} \left(\nabla \varphi\right)^2 + 
\frac{4}{3} \varphi \nabla^2 \varphi \right] 
\delta^i_j + \partial^i \partial_j 
\left(\Phi_{\rm S} - \Psi_{\rm S}\right) 
+ 2 \partial^i \varphi \partial_j \varphi + 4 \varphi \partial^i \partial_j 
\varphi \nonumber \\
&& 
+ \frac{1}{2} \left(\partial^i \dot{V}_j + \partial_j 
\dot{V}^i \right) + {\cal H} \left( \partial^i V_j+ \partial_j 
V^i \right) 
+ \left( \ddot{H}^i_{{\rm S}j} + 2{\cal H} 
\dot{H}^i_{{\rm S}j}
- \nabla^2  H^i_{{\rm S}j}\right)  
= 8 \pi G a^2 
\bar\rho \left( v_{\rm P}^i v_{{\rm P}j} - \frac{1}{3} v_{\rm P}^2 
\delta^i_j \right)  \;. 
\end{eqnarray}

To deal with this equation, we first apply the operator 
$\partial^j \partial_i$, which allows to solve for the 
combination $\Phi_{\rm S} -\Psi_{\rm S}$, and then replace it in  
the original equation, together with the expression for the 
vector mode $V_i$.  
After a lengthy but straightforward calculation we obtain  
\begin{equation}
\label{secondgravwaves}
\nabla^2 \nabla^2
\left( \ddot{H}^i_{{\rm S}j} + 2{\cal H} \dot{H}^i_{{\rm S}j}
- \nabla^2  H^i_{{\rm S}j}\right)  = \nabla^2 \partial^k 
\partial_\ell R^\ell_{~k} \delta^i_j 
+ 2 \nabla^2 \left(\nabla^2 R^i_{~j} 
-  \partial^i \partial_k R^k_{~j} - \partial^k \partial_j 
R^i_{~k} \right) + \partial^i \partial_j \partial^k \partial_\ell R^\ell_{~k} 
\;, 
\end{equation}
where we introduced the traceless tensor  
\begin{eqnarray}
R^\ell_{~k} & \equiv & \partial^\ell \varphi \partial_k \varphi 
- \frac{1}{3} \left( \nabla \varphi \right)^2 \delta^\ell_k 
+ 4 \pi G a^2 \bar \rho \left( v^\ell_{\rm P} v_{{\rm P}k} - 
\frac{1}{3} v_{\rm P}^2 \delta^\ell_k \right) \nonumber \\
& = & g^2 \left(1 + \frac{2E^2(z) f^2(\Omega_m)}{3 \Omega_{0m} 
\left(1+z\right)^3}
\right) \left( \partial^\ell \varphi_0 \partial_k  \varphi_0 
- \frac{1}{3} \left( \nabla \varphi_0 \right)^2 \delta^\ell_k \right)
\;.
\end{eqnarray}

Equation (\ref{secondgravwaves}) can be solved by Green's method, as the 
corresponding homogeneous equation is the one for linear tensor modes, whose 
analytical solutions are known in the limiting cases $\Omega_m \to 1$ or 
$\Omega_\Lambda \to 1$.
In the $\Lambda \to 0$ limit one recovers the result of Ref. \cite{mmb},
namely    
\begin{equation}
\label{eq:tensors}
H_{{\rm S}ij}({\bf x},\eta) = \frac{1}{(2\pi)^3} \int d^3 k
e^{i{\bf k}\cdot{\bf x}}\frac{40}{k^4} {\cal S}_{ij}({\bf k})\left(
\frac{1}{3} - \frac{j_1(k\eta)}{k\eta} \right),
\end{equation}
where $j_\ell$ are spherical Bessel functions of order $\ell$ 
and ${\cal S}_{ij}({\bf k}) = \int d^3 y e^{-i{\bf k}\cdot{\bf y}}
{\cal S}_{ij}({\bf y})$, with 
\begin{equation}
\label{eq:source}
{\cal S}_{ij}=\nabla^2 \Theta_0 \delta_{ij} + \partial_i 
\partial_j \Theta_0 + 2\left( \partial_i \partial_j 
\varphi_0 \nabla^2\varphi_0 - \partial_i \partial_k
\varphi_0 \partial^k \partial_j \varphi_0\right) \;. 
\end{equation} 

For the purpose of the present analysis, the main contribution comes 
from the early evolution of $H_{{\rm S}ij}$, when the cosmological 
constant term can be neglected. One can then adopt the 
Einstein-de Sitter result above, up to an overall correction factor 
$g_\infty^2 \equiv g^2(z\to\infty)$.

It proves convenient to express the previous quantities in Fourier space.
Let us define 
\begin{equation}
\label{eq:vectorexp}
V_j({\bf x},\eta) = 
\frac{1}{(2 \pi)^3} \int d^3 k \left[V^{(+1)}({\bf k},\eta)  
Q^{(+1)}_j({\bf k},{\bf x}) + V^{(-1)}({\bf k},\eta)  
Q^{(-1)}_j({\bf k},{\bf x}) \right] \;, 
\end{equation} 
where \cite{hw} 
\begin{equation}
Q^{(\pm 1)}_j({\bf k},{\bf x}) = \frac{i}{\sqrt 2} ({\hat {\bf e}}_1 \pm 
i {\hat {\bf e}}_2 )_j \exp(i{\bf k} \cdot {\bf x}) \;,
\end{equation}
with ${\hat {\bf e}}_1$, ${\hat {\bf e}}_2$ and ${\hat {\bf e}}_3 = 
{\hat {\bf k}}$ forming an orthonormal basis and 
\begin{equation}
\label{eq:vamplitude}
V^{(\pm 1)}({\bf k},\eta) = \frac{\sqrt 2}{3} \frac{F(z)}{k^2} \int 
\frac{d^3 k^\prime}{(2 \pi)^3} ({\hat {\bf e}}_1 \mp i {\hat {\bf e}}_2) \cdot 
{\bf k}^\prime
\left(k^2 - 2 {\bf k} \cdot {\bf k}^\prime\right) \varphi_0({\bf k}^\prime) 
\varphi_0({\bf k}- {\bf k}^\prime) \;. 
\end{equation}

The peculiar gravitational potential $\varphi_0$ is a Gaussian random 
field with Fourier space correlation 
\begin{equation}
\langle \varphi_0({\bf k}) \varphi_0({\bf k}^\prime) \rangle
= (2 \pi)^3 \delta^{(3)}({\bf k}+{\bf k}^\prime) P_\varphi(k) \;,
\end{equation}
where $P_\varphi(k)$ is the gravitational potential 
power-spectrum $P_\varphi(k)= P_{0\varphi} k^{-3} (k/k_0)^{n_s-1}
T^2_{\rm s}(k)$, $k_0$ is some pivot wave-number and 
$T_{\rm s}$ is the usual matter transfer function \cite{bbks}, 
which is unity on large scales and drops off like $k^{-2}$ on small 
scales because of the stagnation effect of matter perturbations on 
sub-horizon scales during radiation dominance. 
For later convenience we can relate $P_\varphi$
to the dimensionless power-spectrum of the 
comoving curvature perturbation (see e.g. \cite{peiris03}) 
$\cal R$, namely $\Delta^2_{\cal R}(k) = \Delta^2_{\cal R}(k_0) 
(k/k_0)^{n_s-1}$, where $\Delta^2_{\cal R}(k_0) = (25/9) 
(P_{0\varphi}/2\pi^2)$ 
\footnote{The relation with the inflation
parameters is $\Delta^2_{\cal R}=H^2/(\pi\epsilon m_P^2)$, with 
$m_P\equiv G^{-1/2}$}.

The vectors $V^{(\pm 1)}$ are two independent, {\it chi-square} distributed
random fields each one with power-spectrum 
\begin{equation}
P_V(k,\eta) = \frac{1}{9 \pi^2} \frac{F(z)^2}{k^4} \int_{-1}^1 d\cos\theta
(1-\cos^2\theta) 
\int_0^\infty d k^\prime k^{\prime 4} \left(k^2 - 2 k k^\prime \cos \theta
\right)^2 P_\varphi(k^\prime) 
P_\varphi({\sqrt{k^{\prime 2} + k^2 - 2 k^\prime k\cos \theta}}) \;,
\end{equation}
which, after numerical integration yields 
\begin{equation}
P_V(k) = \frac{36\pi^2}{(25)^2} 
C_V(n_s) \Delta^4_{\cal R}(k_0) (k F(z))^2 k^{-3} 
\left(\frac{k}{k_\ast}\right)^{-1} 
\left(\frac{k_\ast}{k_0}\right)^{2(n_s-1)}
W_V\left(k/k_\ast\right) \;,
\label{psv}
\end{equation}
where $k_\ast=\Omega_{0m}h^2$ Mpc$^{-1}$, with 
$h$ the Hubble constant $H_0$ in units of 100 km s$^{-1}$ Mpc$^{-1}$; 
the coefficient $C_V(n_S)$ ranges from $0.062$ to $0.29$ for $n_s$  
between $1$ and $0.8$, respectively. 
The function $W_V(x)$ (which is also weakly dependent on $n_s$) is unity 
for $x \ll 1$ and drops to zero at 
$x \approx 1$; its form is plotted in Figure 1 (for $n_s=1$). 
A good fit, useful for next section calculations, is given by
$W_V(x)=(1+5x+3x^2)^{-5/2}$.
\begin{figure}[t]
\includegraphics[width=10cm]{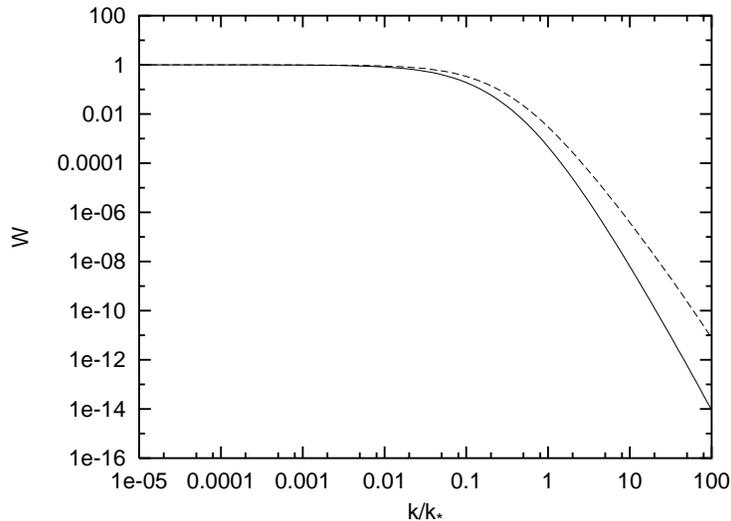}\\
\caption[fig0]{\label{fig1} The functions $W_V$ (dashed line) and $W_\chi$
(solid line) for $n_s=1$.}
\end{figure}

For tensor modes we similarly define
\begin{equation}
\label{eq:tensorexp}
H_{ij}({\bf x},\eta) = \frac{1}{(2 \pi)^3} \int d^3 k 
\left[H^{(+2)}({\bf k},\eta)  
Q^{(+2)}_{ij}({\bf k},{\bf x}) + H^{(-2)}({\bf k},\eta)  
Q^{(-2)}_{ij}({\bf k},{\bf x}) \right] \;, 
\end{equation} 
where \cite{hw}
\begin{equation}
Q^{(\pm 2)}_{ij}({\bf k},{\bf x}) = - {\sqrt \frac{3}{8}} 
({\hat {\bf e}}_1 \pm i {\hat {\bf e}}_2 )_i  ({\hat {\bf e}}_1 \pm i 
{\hat {\bf e}}_2 )_j  \exp(i{\bf k} \cdot {\bf x}) \;.
\end{equation}
The primary tensor modes can be expressed as 
\begin{equation}
\label{eq:primatamplitude}
H^{(\pm 2)}_P({\bf k},\eta) = \chi_P^{(\pm 2)}({\bf k}) 
\frac{3j_1(k\eta)}{k\eta}  \;, 
\end{equation}
for the Einstein-de Sitter case, where $\chi_P^{(+2)}$ and 
$\chi_P^{(-2)}$ are two independent Gaussian fields
each one with power-spectrum 
$P_{\chi_P}(k) = P_{0\chi_P} k^{-3} (k/k_0)^{n_t}T_{\rm t}^2(k)$. 
The tensor transfer function $T_{\rm t}(k)$ reads 
$T_{\rm t}(k) = [1 + 1.34 y + 2.5 y^2]^{1/2}$  
\cite{turwhite}, where $y\equiv k/k_{\rm eq}$ and 
$k_{\rm eq} = 0.073 \Omega_{0m} h^2$ Mpc$^{-1}$ 
is the horizon scale at matter radiation equality. In the $\Lambda \neq 0$ 
case the evolution of primary tensor modes is more complicated and will 
be dealt with using CMBFAST \cite{se96}. 

We can similarly relate
$P_{\chi_P}(k)$ to the dimensionless tensor power-spectrum 
(see e.g. \cite{peiris03}),
$\Delta^2_{\rm t}(k) = \Delta^2_{\rm t}(k_0) (k/k_0)^{n_t}$, where 
$\Delta^2_{\rm t}(k_0) = (P_{0\chi_P}/2\pi^2)$.  
Defining the `tensor-to-scalar' ratio as in Ref. \cite{peiris03}, 
we find
\begin{equation}
\label{eq:ratio}
r\equiv 24 \frac{\Delta^2_{\rm t}(k_0)}{\Delta^2_{\cal R}(k_0)} \;. 
\end{equation}
In single-field slow-roll inflation models the standard consistency relation
$n_t=-r/8$ also applies \cite{consist}. 

The secondary tensor modes are instead characterized by 
\begin{equation}
\label{eq:secondtampli}
H^{(\pm 2)}_S({\bf k},\eta) \approx 
\left(1 - \frac{3 j_1(k\eta)}{k\eta} \right) g_\infty^2 
\chi_S^{(\pm 2)}({\bf k}) \;, 
\end{equation}
where we have introduced the factor $g_\infty^2$ that makes this a
good approximation in the $\Lambda \neq 0$ case for $z>$ few as
discussed above.
The secondary tensors in Eq. (\ref{eq:secondtampli}) are given by
\begin{equation}
\label{eq:secondtampli2}
\chi_S^{(\pm 2)}({\bf k}) = \frac{10}{3{\sqrt 6}~k^2}  
\int \frac{d^3 k^\prime}{(2 \pi)^3} \left[{\bf k}^\prime \cdot 
({\hat {\bf e}}_1 \mp{\hat {\bf e}}_2)\right]^2 
\varphi_0({\bf k}^\prime) \varphi_0({\bf k}- {\bf k}^\prime) \;,
\end{equation}
and are {\it chi-square} distributed independent random fields 
with power-spectra
\begin{equation}
P_{\chi_S}(k) = \frac{25}{27 \pi^2} 
\frac{1}{k^4} \int_{-1}^1 d \cos \theta \sin^4 \theta
\int_0^\infty dk^\prime k^{\prime 6} P_\varphi(k^\prime) 
P_\varphi({\sqrt{k^{\prime 2} + k^2 - 2 k^\prime k \cos \theta}}) \;,
\end{equation}
which can be numerically integrated to give 
\begin{equation}
P_{\chi_S} = \frac{12 \pi^2}{25} C_{\chi_S} \Delta^4_{\cal R}(k_0) k^{-3} 
\left(\frac{k}{k_\ast}\right)^{-1} 
\left(\frac{k_\ast}{k_0}\right)^{2(n_s-1)} 
W_{\chi_S}\left(k/k_\ast\right) \;, 
\label{pschis}
\end{equation} 
where the coefficient $C_{\chi_S}(n_S)$ ranges from 
$0.062$ to $0.29$ for $n_s$ between $1$ and $0.8$, respectively. 
The function $W_{\chi_S}(x)$ (which weakly depends on $n_s$) 
is unity for $x \ll 1$ and drops to zero at 
$x \approx 1$; its form is also plotted in Figure 1 (for $n_s=1$). 
It can be well fitted by $W_\chi(x)=(1+7x+5x^2)^{-3}$.

\section{Polarization angular power-spectra} 

Polarization of the CMB arises from Thomson scattering 
of anisotropic radiation by free electrons. The generation of temperature 
and polarization anisotropies in the CMB from gravitational perturbations 
has been studied in detail by several authors \cite{se97,ka97,hw,hu98}.
A simple and powerful formalism is the total angular momentum method 
\cite{hw,hu98}, which includes the effect of scalar, vector and tensor 
modes on an equal footing. This is the most convenient approach for our 
calculations, and we are going to extensively use the results of Ref. 
\cite{hw}. 
The temperature and polarization fluctuations are expanded in normal modes
that take into account the dependence on both the angular direction of 
photon propagation ${\bf n}$ and the spatial position ${\bf x}$,
$\Spin{G}{s}{\ell}{m}({\bf x},{\bf n})$
\begin{equation}
\begin{array}{rcl}
\Theta(\eta,{\bf x},{\bf n}) &=& \displaystyle{
\int \frac{d^3 k}{(2\pi)^3} }
        \sum_{\ell}
\sum_{m=-2}^2 \Theta_\ell^{(m)} \Gm{0}{\ell}{m} \, , \\
 (Q \pm i U)(\eta,{\bf x},{\bf n}) &=& \displaystyle{\int \frac{d^3k}
 {(2\pi)^3}}
        \sum_{\ell} \sum_{m=-2}^2
        (E_\ell^{(m)} \pm i B_\ell^{(m)}) \, \Gm{\pm 2}{\ell}{m} \,, 
\end{array}
\label{eqn:decomposition}
\end{equation}
with spin $s=0$ describing the temperature fluctuation and $s=\pm 2$
describing the polarization tensor and $m=0, \pm 1, \pm 2$ denoting scalar, 
vector and tensor perturbations, respectively.
$E_\ell^{(m)}$ and $B_\ell^{(m)}$ are the angular moments of the
electric and magnetic polarization components and  
\begin{equation}
\Gm{s}{\ell}{m}({\bf x},{\bf n})  =
        (-i)^\ell \sqrt{ \frac{4\pi}{2\ell+1}}
        [\Spy{s}{\ell}{m}({\bf n})] \exp(i{\bf k} \cdot {\bf x})\,.
\end{equation}

The Boltzmann equation describing the time evolution of the radiation 
distribution under gravitation and scattering processes can be written as a
set of evolution equations for the angular moments of the temperature,
$\Theta_\ell^{(m)}$ (for $\ell \ge m)$, and both polarization types, 
$E_\ell^{(m)}$ and $B_\ell^{(m)}$ (for $\ell \ge 2$ and $m \ge 0$), 
\begin{eqnarray}
\dot \Theta_\ell^{(m)}&=& k
\Bigg[ \frac{\Spin{\kappa}{0}{\ell}{m}}{(2\ell-1)}
        \Theta_{\ell-1}^{(m)}
             -\frac{\Spin{\kappa}{0}{\ell+1}{m}}{(2\ell+3)}
        \Theta_{\ell+1}^{(m)} \Bigg]
	- \dot\tau \Theta_\ell^{(m)} + S_\ell^{(m)}, \\
\dot E_\ell^{(m)}&=& k \Bigg[ {\Spin{\kappa}{2}{\ell}{m} \over (2\ell-1)}
E_{\ell-1}^{(m)} - {2m \over \ell (\ell + 1)} B_\ell^{(m)} 
- {\Spin{\kappa}{2}{\ell+1}{m} \over (2 \ell + 3)}
        E_{\ell + 1}^{(m)} \Bigg] 
	- \dot\tau [E_\ell^{(m)}+\sqrt{6}P^{(m)}\delta_{\ell,2}]\, , \\
\dot B_\ell^{(m)}&=& k \Bigg[ {\Spin{\kappa}{2}{\ell}{m} \over (2\ell-1)}
B_{\ell-1}^{(m)} + {2m \over \ell (\ell + 1)} E_\ell^{(m)} 
- {\Spin{\kappa}{2}{\ell+1}{m} \over (2 \ell + 3)}
        B_{\ell + 1}^{(m)} \Bigg] -\dot\tau B_\ell^{(m)},
\label{eqn:boltzmann}
\end{eqnarray}
where the coupling coefficients are
\begin{equation}
\Spin{\kappa}{s}{\ell}{m} = \sqrt{ 
{(\ell^2-m^2)(\ell^2-s^2)\over\ell^2}}.
\end{equation}
The fluctuation sources are given by
\begin{equation}
\begin{array}{lll}
S_0^{(0)} = \dot\tau \Theta_0^{(0)} -  \dot\Phi \, ,      \qquad &
S_1^{(0)} = \dot\tau v_B^{(0)} + k\Psi \, ,               \qquad &
S_2^{(0)} = \dot\tau P^{(0)} \, , \vertsp\\
                                                \qquad &
S_1^{(1)} = \dot\tau v_B^{(1)} + \dot V \, ,              \qquad &
S_2^{(1)} = \dot\tau P^{(1)} \, , \vertsp\\
                                                \qquad &
                                                \qquad &
S_2^{(2)} = \dot\tau P^{(2)} - \dot H      \vertsp \, ,
\label{eqn:S}
\end{array}
\end{equation}
with
\begin{equation}
P^{(m)} = {1 \over 10} \left[ \Theta_2^{(m)}  -
{\sqrt 6} E_2^{(m)} \right]. 
\label{eqn:polsource}
\end{equation}
The modes with $m=-|m|$ satisfy the same equations with  $B_\ell^{(-|m|)}= 
-B_\ell^{(|m|)}$ and all the other quantities unchanged.

These equations can be formally integrated, leading to  
simple expressions in terms of an integral along the line-of-sight \cite{hw}. 
The temperature fluctuations are given by
\begin{equation}
{\Theta_\ell^{(m)}(\eta,k) \over 2\ell + 1}\, 
  =  
\int_0^{\eta} d\eta' \, e^{-\tau} \sum_{\ell'} \,
  S_{\ell'}^{(m)}(\eta') \, j_\ell^{(\ell'm)}(k(\eta-\eta')) \, ,
\label{eqn:inttemp}
\end{equation}
where $j_\ell^{(\ell'm)}$ are given in ref. \cite{hw}.
For the polarization, we have
\begin{eqnarray}
{E^{(m)}_\ell(\eta_0,k) \over 2\ell+1} &=&  -\sqrt{6}
\int_0^{\eta_0} d\eta \, \dot\tau e^{-\tau} P^{(m)}_{\vphantom{\ell}}
(\eta)\epsilon_\ell^{(m)}(k(\eta_0-\eta)) \, ,\nonumber\\
{B^{(m)}_\ell(\eta_0,k) \over 2\ell+1} &=&  -\sqrt{6} 
\int_0^{\eta_0} d\eta \, \dot\tau e^{-\tau} P^{(m)}_{\vphantom{\ell}}
(\eta)\beta_\ell^{(m)}(k(\eta_0-\eta)) \, ,
\label{eqn:pol}
\end{eqnarray}
where the radial functions read
\begin{eqnarray}
\epsilon^{(\pm 1)}_\ell(x) &=& {1 \over 2} {\sqrt {(\ell-1)(\ell+2)}}
        \left[ {j_\ell(x) \over  x^2} + {j_\ell'(x) \over  x}
        \right] \, , \nonumber\\ 
\epsilon^{(\pm 2)}_\ell(x) &=& {1 \over 4} \left[ -j_\ell(x)
        + j_\ell''(x) + 2{j_\ell(x) \over x^2} +
        4{j_\ell'(x) \over x} \right]  \, , \\
\beta^{(+1)}_\ell(x) & = & - \beta^{(-1)}_\ell(x) = 
{1 \over 2} {\sqrt {(\ell-1)(\ell+2)}}~
        {j_\ell(x) \over x} \, , \nonumber\\
\beta^{(+2)}_\ell(x) &=& -\beta^{(-2)}_\ell(x) = 
{1 \over 2} \left[ j_\ell'(x)
        + 2 {j_\ell(x) \over x} \right] \, .
\end{eqnarray}
Here the differential optical depth $\dot \tau = n_e \sigma_T a$ sets 
the collision rate in conformal time, with $n_e$ the free electron density
and $\sigma_T$ the Thomson cross section and
$\tau(\eta_0,\eta) \equiv \int_\eta^{\eta_0} \dot\tau(\eta') 
d\eta^\prime$ 
the optical depth between $\eta$ and the present time. 
The combination $\dot\tau e^{-\tau}$ is the {\em visibility function}
and expresses the probability that a photon last scattered between
$d\eta$ of $\eta$ and hence is sharply peaked at the last scattering
epoch. In early reionization models, a second peak is also present at
more recent times.  

Scalar modes do not contribute to B-polarization, thus
$B^{(0)}_\ell=0$. 
We are interested in the contribution to the angular power-spectrum 
for the E and B modes arising from vector ($m=1$) and tensor ($m=2$)
perturbations, 
\begin{eqnarray}
C_\ell^{(E)\hphantom{E}} 
		  & = &{2 \over \pi} \int {dk \over k}
	\sum_{m=-2}^2
        k^3 \frac{|E_\ell^{(m)}(\eta_0,k)|^2}{(2\ell+1)^2} , \nonumber\\
C_\ell^{(B)\hphantom{E}} 
		  & = &{2 \over \pi} \int {dk \over k}
	\sum_{m=-2}^2
        k^3 \frac{|B_\ell^{(m)}(\eta_0,k)|^2}{(2\ell+1)^2}. 
\end{eqnarray}

\subsection{Effects of decoupling}

Before recombination, photons, electrons and baryons behave as a
single tightly coupled fluid, so we can find approximate expressions
for the polarization source $P^{(m)}$ using a perturbative expansion
in inverse powers of the differential optical depth $\dot\tau$
($k/\dot\tau\ll 1$) \cite{zh95}. 
For the vector modes we need to consider the Euler equation for the
velocity perturbation of baryons to close the system
\begin{equation}
 \dot v^{(1)}_B = \dot V  -  {\dot a \over a}
 (v^{(1)}_B - V) + {\dot\tau \over R} (\Theta_1^{(1)} - v^{(1)}_B) \, .
\end{equation}
To leading order in the tight coupling  approximation we obtain
\begin{equation}
\Theta_1^{(1)}=v_B^{(1)},
\end{equation} 
and from the baryon Euler equation
\begin{equation}
v_B^{(1)}\simeq V.
\end{equation} 
To this perturbative order the quadrupole vanishes. However it does not
vanish to the next order, and thus also $P^{(1)}\neq 0$.
An appropriate expression for our calculation can be obtained by
combining  the equations for $\Theta_2^{(1)}$ and $E_2^{(1)}$ to
obtain an evolution equation for $P^{(1)}$, and replacing in it the 
leading order solution we get
\begin{equation}
\dot P^{(1)}+\frac{3}{10} \dot \tau P^{(1)}-\frac{k}{10{\sqrt 3}}V=0.
\end{equation} 
This can be integrated to give 
\begin{equation} 
P^{(1)}(k,\eta)=\frac{k}{10{\sqrt 3}}\int_0^\eta {\rm d}\eta' V(k,\eta')
\exp(-\frac{3}{10}\tau(\eta',\eta)).
\end{equation} 
Then, the B-mode polarization multipoles can be estimated by 
\begin{equation}
\frac{B_\ell^{(1)}(\eta_0,k)}{2\ell+1}=-\frac{k{\sqrt 2}}{10}
\int_0^{\eta_0} {\rm d}\eta \dot \tau e^{-\tau}
\beta_\ell^{(1)}(k(\eta_0-\eta))\int_0^\eta {\rm d}\eta' V(k,\eta')
\exp\left(-\frac{3}{10}\tau(\eta',\eta)\right),
\end{equation} 
An analogous expression holds for $E_\ell^{(1)}$ replacing
$\beta_\ell^{(1)}$ by $\epsilon_\ell^{(1)}$.
Because of the sharpness of the visibility function around the
time of decoupling $\eta_D$, $V(k,\eta')$ and 
$\beta_\ell^{(1)}(k(\eta_0-\eta))$ (or $\epsilon_\ell^{(1)}$) can be
evaluated at $\eta_D$ and taken out of the integration. The
remaining integral can be evaluated analytically approximating the
visibility function by a Gaussian of width $\Delta\eta_D$ \cite{zh95}
\begin{equation}
\frac{B_\ell^{(1)}(\eta_0,k)}{2\ell+1}=-\frac{k{\sqrt 2}}{3}
0.51 \Delta\eta_D V(k,\eta_D) \beta_\ell^{(1)}(k(\eta_0-\eta_D)).
\end{equation} 
Then, the resulting angular power-spectrum of B modes from vector
perturbations results
\begin{equation}
C_\ell^{(B)(1)} = {8 \over 9\pi} 0.51^2 \Delta\eta_D^2 \int {\rm d}k\ k^4
\left[\beta_\ell^{(1)}\left(k(\eta_0-\eta_D)\right)\right]^2 P_V(k,\eta_D),
\end{equation} 
where we have added the contribution coming from the $m=\pm 1$ modes.
For $C_\ell^{(E)(1)}$ we obtain an analogous expression with 
$\epsilon_\ell^{(1)}(k(\eta_0-\eta_D))$ replacing
$\beta_\ell^{(1)}(k(\eta_0-\eta_D))$.
Replacing Eq. (\ref{psv}) we obtain
\begin{eqnarray}
\frac{\ell (\ell+1) C_\ell^{(B)(1)}}{2\pi}=
\ell (\ell+1) & 16 \frac{0.51^2}{25^2} \Delta^4_{\cal R}(k_0) 
\left(\frac{k_\ast}{k_0}\right)^{2(n_s-1)}C_V(n_s)
\left(\frac{\Delta \eta_D}{\eta_0}\right)^2
\left(\frac{F(z_D)}{\eta_0}\right)^2\eta_0k_\ast \\
&\times
\int {\rm d}x x^2(\beta_\ell^{(1)}(x))^2 
W_V\left(\frac{x}{\eta_0k_\ast}\right).\nonumber
\end{eqnarray} 
The integral can be conveniently evaluated using the WKB
approximation for the Bessel functions appearing in 
$\beta_\ell^{(1)}$ and $\epsilon_\ell^{(1)}$, leading to
\begin{eqnarray}
\langle (\beta_\ell^{(1)})^2(x) \rangle
& \simeq & \frac{1}{8} (\ell-1)(\ell+2)
\frac{1}{x^3{\sqrt {x^2-\ell^2}}}\\
\langle (\epsilon_\ell^{(1)})^2(x) \rangle
& \simeq & \frac{1}{8} (\ell-1)(\ell+2)
\left(\frac{1}{x^5{\sqrt {x^2-\ell^2}}}+\frac{{\sqrt {x^2-\ell^2}}}{x^5}
\right),
\end{eqnarray}
for $x>\ell$, and 0 for $x<\ell$.
The results for the vector contribution to the B and E angular 
power-spectrum are shown in Figure 2, where we have used the following 
values for the  parameters, for the concordance $\Lambda$CDM model 
($\Omega_\Lambda=0.7$, $\Omega_m=0.3$, $\Omega_bh^2=0.022$, $h=0.71$): 
$\Delta \eta_D/\eta_D=\Delta z_D/2z_D=0.09$, $\eta_0/\eta_D=30$,
$k_\ast=0.135$ Mpc$^{-1}$ and $\Delta^2_{\cal R}(k_0)=2.3\times10^{-9}$
at $k_0=0.002$ Mpc$^{-1}$.

For the tensor modes, the tight coupling approximation to the leading
order gives $P^{(2)}=\Theta^{(2)}_2/4=-\dot H^{(2)}/3\dot\tau$. To the 
next order we obtain the following evolution equation for $P^{(2)}$
\begin{equation}
\dot P^{(2)}+\frac{3}{10} \dot\tau P^{(2)}+\frac{\dot H^{(2)}}{10}=0,
\end{equation} 
that can be integrated
\begin{equation}
P^{(2)}(k,\eta)=-\frac{1}{10}\int_0^\eta {\rm d}\eta' 
\dot H^{(2)}(k,\eta')
\exp(-\frac{3}{10}\tau(\eta',\eta)).
\end{equation} 
Proceeding as for the vector case, we obtain for the B modes
\begin{equation}
\frac{B_\ell^{(2)}(\eta_0,k)}{2\ell+1}={\sqrt \frac{2}{3}}
0.51 \Delta\eta_D \dot H^{(2)}(k,\eta_D) 
\beta_\ell^{(2)}(k(\eta_0-\eta_D)).
\end{equation} 
and the same result changing $\beta_\ell^{(2)}$ by
$\epsilon_\ell^{(2)}$ for the E modes.

Using Eq. (\ref{eq:secondtampli}) we can write
\begin{equation}
\dot H^{(2)}_S(k,\eta_D) = -3 g_\infty^2 {j_2(k\eta_D)\over\eta_D} 
\chi^{(2)}_S(k).
\end{equation} 
Then, for the angular power-spectrum, we obtain 
\begin{equation}
\frac{\ell (\ell+1) C_\ell^{(B)(2)}}{2\pi}=
\ell (\ell+1){12 \over \pi^2} 0.51^2 
\left(\frac{\Delta\eta_D}{\eta_D}\right)^2 g_\infty^4
\int {\rm d}k\ k^2
\left[ \beta_\ell^{(2)}(k(\eta_0-\eta_D))\right]^2  P_{\chi_S}(k)
j_2^2(k\eta_D), 
\label{clb2s}
\end{equation} 
for the B modes (and the same expression with
$\beta_\ell^{(2)} \rightarrow 
\epsilon_\ell^{(2)}$ for the E modes), with $P_{\chi_S}(k)$ given by
Eq. (\ref{pschis}).
To perform the remaining integration it is convenient to use the WKB
approximated expressions
\begin{eqnarray}
\langle (\beta_\ell^{(2)})^2(x) \rangle
& \simeq & \frac{1}{8} 
\left(\frac{{\sqrt {x^2-\ell^2}}}{x^3}+
\frac{4}{x^3{\sqrt {x^2-\ell^2}}}\right)\\
\langle (\epsilon_\ell^{(2)})^2(x) \rangle
& \simeq & \frac{1}{8} 
\left[\left(1-\frac{1+\ell(\ell+1)/2}{x^2}\right)^2
\frac{1}{x{\sqrt {x^2-\ell^2}}}+\frac{{\sqrt {x^2-\ell^2}}}{x^5}\right],
\nonumber
\label{be2}
\end{eqnarray}
for $x>\ell$, and 0 for $x<\ell$.
The results are shown in Figure 2.
The dominant contribution comes from the tensor modes for low
multipoles, $\ell<20$, while the vector contribution dominates for
larger $\ell$.

\subsection{Effects of reionization}

The recent detection by {\it WMAP} 
of excess power on large angular scales in the 
correlation between temperature and E-polarization of the CMB is a strong 
evidence that the Universe was reionized at relatively early times 
\cite{bennett03,kogut03}. The best fit to the Thomson scattering optical depth 
due to reionization is given by $\tau_{\rm ri}=0.17$, with still large 
uncertainties. In a spatially-flat $\Lambda$CDM model with 
$\Omega_\Lambda=0.7$, reionization would have happened at a redshift 
$z_{\rm ri}=17$, if it took place in a single step.
 
Early reionization provides another opportunity for Thomson scattering to 
modify the polarization properties of the CMB. Its effects can be 
approximately decomposed in two parts \cite{za97}. 
On the one hand, rescattering 
damps already present anisotropies and polarization by a factor 
$e^{-\tau_{\rm ri}}$. On the other hand, rescattering of the existing 
quadrupole anisotropy significantly enhances the polarization signal at 
low multipoles. 

Here we estimate the effects of reionization on the polarization signal 
induced by secondary vector and tensor modes.
We shall assume, for simplicity, that reionization took place in a single 
step. 
We can then approximate in Eq. (\ref{eqn:pol}) that the visibility function 
$\dot\tau e^{-\tau}$ is sharply peaked at $\eta_{\rm ri}$ and thus evaluate
$\beta_\ell^{(m)}(k(\eta_0-\eta))$ (and $\epsilon_\ell^{(m)}$) at 
$\eta=\eta_{\rm ri}$ and take them out of the integral. 
The coupling between 
photons and electrons after reionization is, however, not very tight, and 
thus we can not proceed with the same approximations as in the previous 
section to compute the polarization sources $P^{(m)}(\eta)$. Indeed, 
$\dot\tau=-\sigma_T n_e a$ with $n_e=0.88 n_{b0} (1+z)^3 X(z)$ and
$X(z)$ the ionization fraction. Thus $\dot\tau=0.0019 H_0 (1+z)^2 X(z) 
(\Omega_b h^2/0.022)(0.71/h)$, and for the cosmological parameters under 
consideration, $k/\dot\tau\approx 0.5 k\eta_0$ at $z=z_{\rm ri}$, and the 
tight coupling approximation is not applicable at any relevant wavelength.

The polarization produced after reionization is essentially due to the 
scattering of existing quadrupole anisotropies, generated by scattering at 
earlier times or by gravitational effects during propagation. Since the 
quadrupole polarization $E_2$ generated during decoupling is typically 
smaller than the temperature quadrupole $\Theta_2$, we can approximate the 
polarization source as $P^{(m)}\approx \Theta^{(m)}_2/10$, and use for 
$\Theta_2^{(m)}$ the formal solution of Eq.~(\ref{eqn:inttemp}), dropping in 
the fluctuation sources $S$ of Eq.~(\ref{eqn:S}) the terms proportional to 
$\dot\tau$ against $\dot V$ (in the case of vectors) or against $\dot H$ 
(for the tensors), since the coupling is not tight. 

The contribution of secondary vectors to B-polarization after reionization 
is then given by
\begin{equation}
\frac{B_\ell^{(1)}(\eta_0,k)}{2\ell+1}=-\frac{\sqrt 6}{2}
\int_{\eta_{\rm ri}}^{\eta_0} {\rm d}\eta \dot \tau e^{-\tau}
\beta_\ell^{(1)}(k(\eta_0-\eta))\int_0^\eta {\rm d}\eta' 
\exp(-\tau(\eta',\eta)) \dot V(k,\eta') j_2^{(11)}(k(\eta-\eta')),
\end{equation} 
where $j_2^{(11)}(x) = {\sqrt 3}j_2(x)/x$. For a single step
reionization scenario, we can approximate
\begin{equation}
\frac{B_\ell^{(1)}(\eta_0,k)}{2\ell+1}=-\frac{3}{\sqrt 2}
\Delta\eta_{\rm ri}\dot V(k,\eta_{\rm ri})
\beta_\ell^{(1)}(k(\eta_0-\eta_{\rm ri}))\frac{1}{k\eta_{\rm ri}}
\left(\frac{j_1(k\eta_{\rm ri})}{k\eta_{\rm ri}}-
\frac{1}{3}\right).
\end{equation} 
We have defined an effective``reionization width'' $\Delta\eta_{\rm ri}
\equiv\eta_{\rm ri}\int_{\eta_{\rm ri}}^{\eta_0} {\rm d}\eta 
\dot \tau e^{-\tau}$, which for the $\Omega_m=0.3$, spatially-flat 
$\Lambda$CDM model with
single-step reionization and optical depth $\tau_{\rm ri}=0.17$ is
$\Delta\eta_{\rm ri}=0.16\eta_{\rm ri}$. This can be calculated taking 
into account that, for $z\le z_{\rm ri}$,
\begin{equation}
\tau(z)=0.0042 \frac{\Omega_b h^2}{0.022}\frac{0.71}{h}\frac{0.3}{\Omega_m}
\left( \sqrt{1-\Omega_m + \Omega_m(1+z)^3}-1\right) 
\end{equation}
Notice that $\tau_{\rm ri}=0.17$ for $z_{\rm ri}=17$ if $\Omega_m=0.3$.
The reionization contribution to the power-spectrum of B-polarization by 
vector modes is finally approximately given by
\begin{eqnarray}
\frac{\ell (\ell+1) C_\ell^{(B)(1)}}{2\pi}=
\ell (\ell+1) & \left(\frac{18}{25}\right)^2 \Delta^4_{\cal R}(k_0) 
\left(\frac{k_\ast}{k_0}\right)^{2(n_s-1)}C_V(n_s)
\left(\frac{\Delta \eta_{\rm ri}}{\eta_{\rm ri}}\right)^2
\left(\frac{F(z_{\rm ri})}{\eta_{\rm ri}}\right)^2 
(\eta_0-\eta_{\rm ri})k_\ast\\
& \int\frac{{\rm d}x}{x^2}(\beta_\ell^{(1)}(x))^2 
\left(\frac{j_1(y)}{y}-\frac{1}{3}\right)^2
W_V\left(\frac{x}{(\eta_0-\eta_{\rm ri})k_\ast}\right).\nonumber
\end{eqnarray} 
where $y=x\eta_{\rm ri}/(\eta_0-\eta_{\rm ri})$. An analogous expression
holds of course for $C_\ell^{(E)(1)}$ with $\epsilon_\ell^{(1)}$ in place of 
$\beta_\ell^{(1)}$. 

The contribution of secondary tensor modes to the polarization induced 
after reionization can be estimated with similar approximations:
\begin{equation}
\frac{B_\ell^{(2)}(\eta_0,k)}{2\ell+1}=-\frac{\sqrt 6}{2}
\int_{\eta_{\rm ri}}^{\eta_0} {\rm d}\eta \dot \tau e^{-\tau}
\beta_\ell^{(2)}(k(\eta_0-\eta))\int_0^\eta {\rm d}\eta' 
\exp(-\tau(\eta',\eta)) \dot H^{(2)}(k,\eta') j_2^{(22)}(k(\eta-\eta')),
\end{equation} 
where $j_2^{(22)}(x) = 3\ j_2(x)/x^2$. We approximate 
\begin{equation}
\frac{B_\ell^{(2)}(\eta_0,k)}{2\ell+1}=-\frac{\sqrt 6}{10}
\Delta\eta_{\rm ri}(1+\delta)
\dot H^{(2)}(k,\eta_{\rm ri})
\beta_\ell^{(2)}(k(\eta_0-\eta_{\rm ri}))D(k\eta_{\rm ri})\ .
\end{equation} 
We have defined an additional coefficient $\delta$ such that 
$\delta\Delta\eta_{\rm ri}\equiv\int_{\eta_{\rm ri}}^{\eta_0} {\rm d}\eta 
\dot \tau e^{-\tau} \int_{\eta_{\rm ri}}^{\eta} {\rm d}\eta' 
e^{-\tau(\eta,\eta')}$. It accounts for the fact that in this case 
(opposite to the vector case) the time dependence of the source after 
reionization is not negligible. For the model parameters used throughout this 
work to estimate the effects, $\delta\approx 0.4$.
The damping factor $D(x)\equiv 4/x$ if $x\le 4$ (and $D=1$ otherwise) 
arises because $j_2^{(22)}(x)$ can be approximated as 1/5 only for small 
$x$; larger values lead
to oscillations that make the integral decay as $1/k$. Finally, the 
reionization contribution to the power-spectrum of B-polarization by 
secondary tensor modes is approximately given by
\begin{equation}
\frac{\ell (\ell+1) C_\ell^{(B)(2)}}{2\pi}=
\ell (\ell+1)  \frac{27}{25\pi^2} 
\left(\frac{\Delta \eta_{\rm ri}}{\eta_{\rm ri}}\right)^2
(1+\delta)^2 g_\infty^4\int {\rm d}k\ k^2
\left[ \beta_\ell^{(2)}(k(\eta_0-\eta_{\rm ri}))\right]^2  P_{\chi_S}(k)
j_2^2(k\eta_{\rm ri})D^2(k\eta_{\rm ri})\ .
\end{equation} 

The numerical result for the angular power-spectrum of E and B polarization 
by secondary vector and tensor modes produced after reionization, in the 
concordance $\Lambda$CDM model, is displayed in Fig. 2, added to the 
polarization that was produced during decoupling, damped by a factor 
$e^{-2\tau_{\rm ri}}$.

\begin{figure}[t]
\includegraphics[width=17cm]{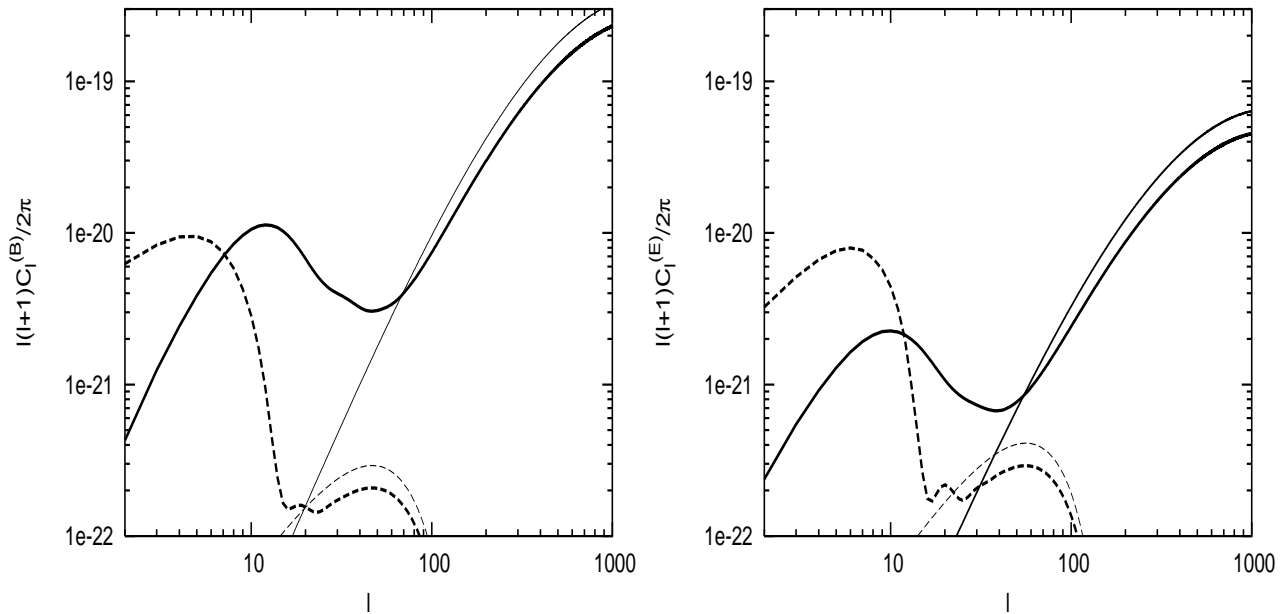}\\
\caption[fig0]{\label{fig2} Angular power-spectrum of 
B (left panel) and E (right panel) polarization by secondary vector 
(solid lines) and tensor modes (dashed lines), in a $\Lambda$CDM model 
without reionization (thin lines) and 
with an optical depth to reionization $\tau_{\rm ri}=0.17$ (thick lines).}
\end{figure}

\section{Comparison with primary tensor modes and cosmic shear}

The signal from the secondary modes computed in the last section
must be compared with the signal from the primordial gravitational-wave
background generated during inflation. This depends on the model of
inflation considered and its amplitude is parameterized by the
'tensor-to-scalar' ratio $r$ (see Eq. (\ref{eq:ratio})). Its angular 
power-spectrum can be accurately computed by available codes like CMBFAST
\cite{se96}, against which we have tested the accuracy of our semi-analytic 
approximations in the previous sections. The polarization induced by 
primordial gravitational waves can be calculated as in Eq.~(\ref{clb2s}) 
replacing the power-spectrum of secondary modes $P_{\chi_S}(k)$ by that of 
the primary modes $P_{\chi_P}(k)$. The approximate expressions, both with or 
without reionization, are in good agreement (with more than 20\% accuracy) 
with the numerical results obtained using CMBFAST. 
Notice that the polarization induced by primordial gravitational waves 
may be slightly smaller than it was calculated here due to damping by 
neutrino free-streaming \cite{weinberg03},
that we did not take into account.

In Figure 3 we plot the numerical result (calculated with CMBFAST) for the 
B-polarization induced by primordial gravitational waves in a model with 
tensor to scalar ratio $r=10^{-6}$, 
along with the total secondary B-modes (vector plus tensor) derived in this 
work.
It is clear that the effects of primary and secondary modes are comparable.
Primary modes have somewhat larger effects at the lowest ($\ell <10$) 
multipoles, and also for $\ell$ between 30 and 100, but secondary modes 
overcome them in intermediate and larger values of $\ell$. The power-spectrum 
induced by primordial gravitational waves scales as $r$. Thus, secondary 
vector and tensor fluctuations limit the ability to detect primordial
gravitational waves through measurements of B-polarization if $r<10^{-6}$. 

The extraction of the B-polarization signature of primordial gravitational 
waves is seriously limited by weak gravitational lensing effects, that 
convert the dominant E-type polarization into B-modes \cite{zalsel98}, 
which swamp the signature of primordial tensor fluctuations unless they have 
a ratio to scalar fluctuations $r$ of the order or larger than
$10^{-4}$. Fortunately, the weak lensing effect can be accounted for, through 
reconstruction of the gravitational lensing potential by the correlations of
B-polarization between large and small angular scales, which primordial 
gravitational waves do not produce. ``Cleaning'' of the gravitational lensing 
signature may achieve factors of 40 in the power-spectrum, or even larger 
\cite{seljak03}.
In Figure 3 we display the predicted gravitational lensing signal in 
B-polarization calculated with CMBFAST for the cosmological model under 
consideration, reduced by a factor 40. Clearly, an improvement in the 
cleaning algorithm would convert the secondary vector and tensor modes 
derived here in the barrier to the detection of
primordial gravitational waves through B-polarization of the CMB if 
$r<10^{-6}$.

\begin{figure}[t]
\includegraphics[width=12cm]{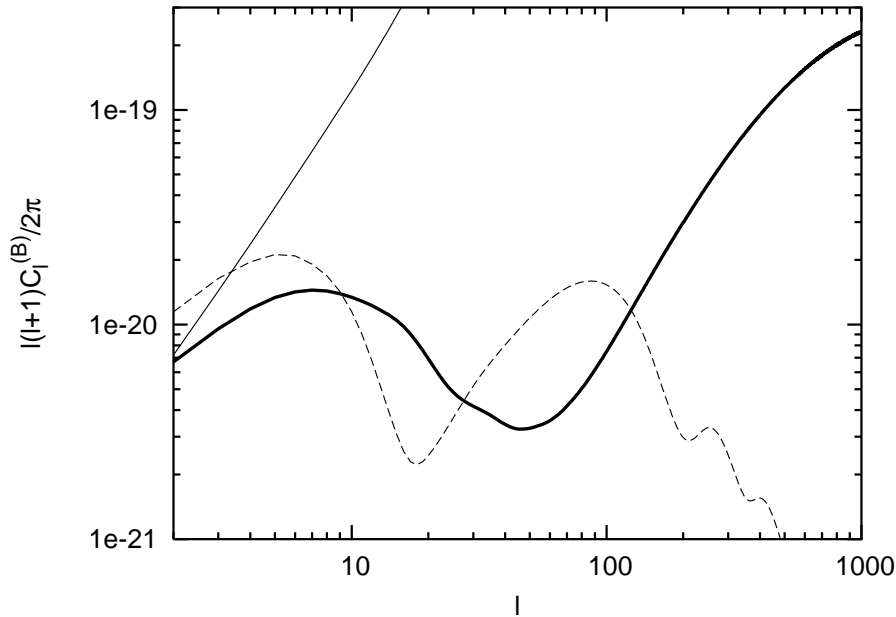}\\
\caption[fig0]{\label{fig3} Angular power-spectrum of secondary total 
(vector plus tensor)
B-polarization (thick solid line) compared with that induced by primordial 
gravitational waves (dashed line) with a tensor to scalar ratio $r=10^{-6}$, 
in a flat $\Lambda$CDM model ($\Omega_m=0.3$) with $\tau_{\rm ri}=0.17$. 
The thin solid line is the signal due to  gravitational lensing 
cleaned by a factor 40.}
\end{figure}

\section{Conclusions}

The study of the magnetic-mode polarization of the CMB will become a 
fundamental and possibly unique tool to search for the stochastic 
gravitational-wave background, whose detection would represent a clear
signature of a period of inflation in the early Universe.  
In contrast with temperature anisotropies and E-type polarization of the CMB,
scalar perturbations do not give rise to B polarization in a direct way, 
so that measurements of the B-mode could be used to probe rather small 
gravitational-wave background amplitudes. 
Because of this reason, much observational effort is taking place for its 
detection, and future dedicated missions, such as NASA's {\it Beyond Einstein}
Inflation Probe \cite{IP} or  
ground-based experiments, like {\it BICEP} \cite{BICEP} and {\it PolarBeaR},  
are being planned.  

The main background for the detection of the B-mode is represented 
by the gravitational lensing conversion of a fraction of the dominant E-type 
into B-type polarization. However, it has been recently shown \cite{seljak03} 
that the lensing signal can largely be cleaned, thus allowing to probe 
inflationary models with tensor-to-scalar ratios $r \le 10^{-6}$. 
As the largest signal compared to the lensing one comes from low multipoles, 
where the reionization contribution is dominant, this limit depends on the 
Thomson scattering optical depth to reionization, that still has a large 
uncertainty. 

At these sensitivity levels, there are other secondary effects that can give
rise to sizable contributions to the B-type polarization. We estimated here 
the contribution coming from secondary vector and tensor modes, which 
originate by the mildly non-linear evolution of primordial density 
perturbations. The amplitude and harmonic content of this contribution is 
completely fixed, once the primordial power-spectrum of the density 
perturbations is known. For a concordance $\Lambda$CDM model, and adopting  
the high reionization redshift implied by the {\it WMAP} data 
\cite{bennett03,kogut03}, we found that the effect of secondary vectors and 
tensors becomes comparable to 
that of primary gravitational waves for $r \le 10^{-6}$, 
thus making it the actual background for the detection of primordial 
gravitational waves through B-mode polarization.

\acknowledgments{Work partially supported by ANPCyT, Fundaci\'on Antorchas
and MIUR. We thank Paolo Natoli for discussions.}

\end{document}